# Comparing the Extinction and Night Sky Background in UBV on La Silla, and on the ALMA site: Preliminary results

E. Giraud[1], G. Vasileiadis[1], P. Valvin[2], and I. Toledo[3]

[1] LPTA Report - Laboratoire Physique Théorique et Astroparticules, UMR5207 In2p3/UM II, F-34095 Montpellier
e-mail: edmond.giraud@lpta.in2p3.fr; georges.vasileiadis@lpta.in2p3.fr
[2] Groupe d'Etude des Semi-Conducteurs, UMR 5650 CNRS-Université Montpellier II F-34095 Montpellier
e-mail: pierre.valvin@ges.univ-montp2.fr
[3] Department of Astronomy and Astrophysics, P. Universidad Catolica de Chile, Casilla 306, Santiago, Chile

**Abstract.** We report on measurements of the extinction in the U, B & V bands and of the NSB (Night Sky Background) during 2 dark periods on La Silla Observatory and at 4000-5000m on the ALMA site using an UV optimized 25 cm portable telescope. Using Landolt stars we obtained the color equations of our filters in the range of $-0.22 < B-V < 1.45$, $-1.11 < U-B < 1.18$, the extinction coefficients within $1 < Z < 2$ air masses, and the NSB in U, B & V on La Silla. We measured the transparency of the night sky and the NSB in U, at 4000m asl. on the ALMA road towards the Chajnantor site, and the extinction in U, B & V at 5000m on the Chajnantor plateau. We also obtained some NSB measurements at 5000m in U under variable sky conditions.

## 1. Introduction

Cherenkov Astronomy measures the energy of electromagnetic showers, produced by high energy photons entering the atmosphere and emitting Cherenkov radiation. This emission is characterized by a short formation time, low flux and a spectrum that is peaked in the UV region. Thus its detection demands a high quality photometric sky, high transparency in the UV, and low NSB. Due to the origin of these showers, the upper part of the atmosphere, the detected flux depends on the altitude.

Searching for a site for Cherenkov Astronomy resembles then similar studies for the Optical one, but more oriented towards good UV transparency and high altitude.

Our project is centered on the characterization of sites of medium to high altitude, oriented towards scientific projects opted for clear skies in the UV. On the other hand, due to limited resources, these studies should not been seen as a comprehensive site study but more as an experimental indication of the quality of the night sky in the UV, that could be used later for more detailed monitoring. In other words our studies should be interpreted as a preparation for a larger project that looks for a photometric site of low NSB, high transparency and well defined atmospheric conditions.

We have observed the sky of a well known site La Silla, that will serve as a comparison for other sites. Then we performed some observations at higher altitude, on two spots at ~ 4000m, and ~ 5000m respectively.

*Send offprint requests to*: E. Giraud

## 2. Telescope and Observations

Our Telescope is a Newton 25 cm equipped with a CCD camera at f/4. It can be disassembled in three parts : the tripod, the mounting and the tube itself, all parts able to be transported by a single person. All necessary equipment is stored in two crates transportable by two persons.

The tube was built by Optical Guidance Systems. Its mirror is coated with AlSiO2. The camera is a SBIG ST7X model cooled by Peltier effect. The field of view of the CCD is 22 arcmin x 15 arcmin giving a scale of 1.7 arcsec/pixel, a figure obtained by measuring standard stars (pixels of $9\mu m$). Quantum efficiency varies between 45 % at 400 nm to 75 % at 600 nm. The tube is mounted on a Takahashi motorized equatorial mount. Guiding is achieved by means of a second CCD placed adjacent to the acquisition one. The whole setup can be run regulated either in 220VAC or 12V batteries. The camera is equipped with a filter wheel, where we have mounted a set of UBVR Bessel filters. A single portable computer is used to command all parts of the system while data are stored on CDs.

Data were collected during 6 nights on La Silla, (26 November - 1 December 2005), 4 nights (4-5 December 2005, and 24-25 March 2006) at 4000m, and 4 nights (27-30 March 2006) at 5000m, along the construction road that leads from the main installation buildings at 3000m of the Atacama Large Millimeter Array (ALMA) project, to the Chajnantor plateau, where the Array will be installed at 5100m. At La Silla we were located outside the Marly Telescope (Fig. 1).

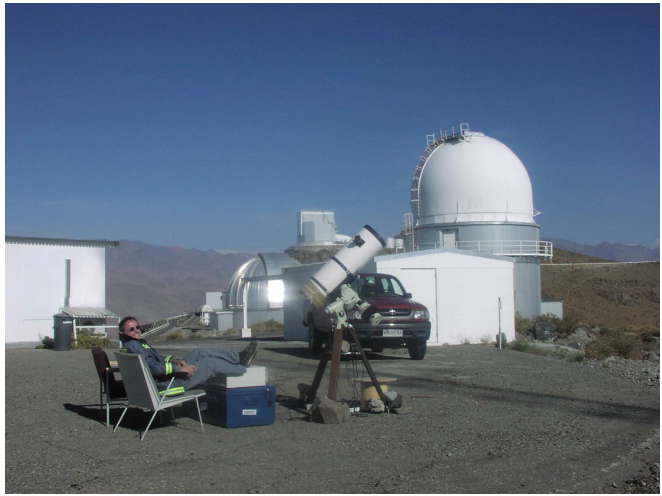

**Fig. 1.** Telescope installation at La Silla just outside the Marly

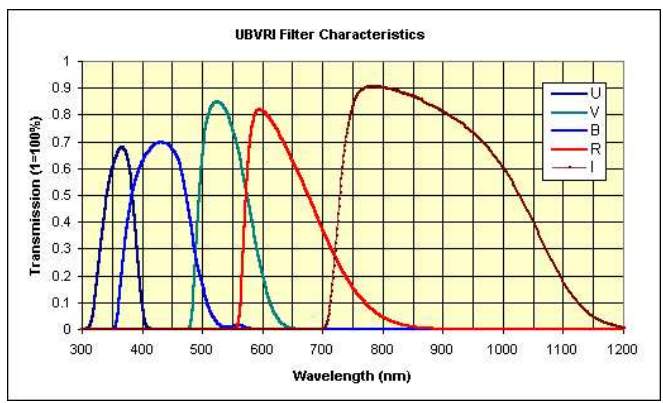

**Fig. 2.** Response curves of the filters used

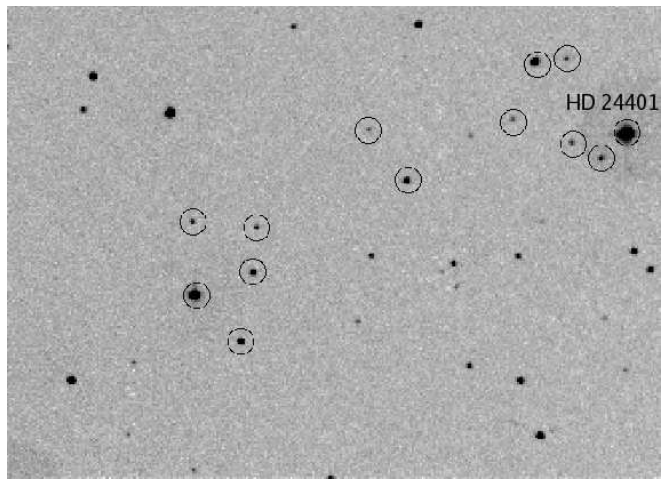

**Fig. 3.** U band image of the field HD 24401. The Landolt standard stars are marked with a circle. Exposure time: 1200 s, scale: 3.4 arcsec/pixel. North is up

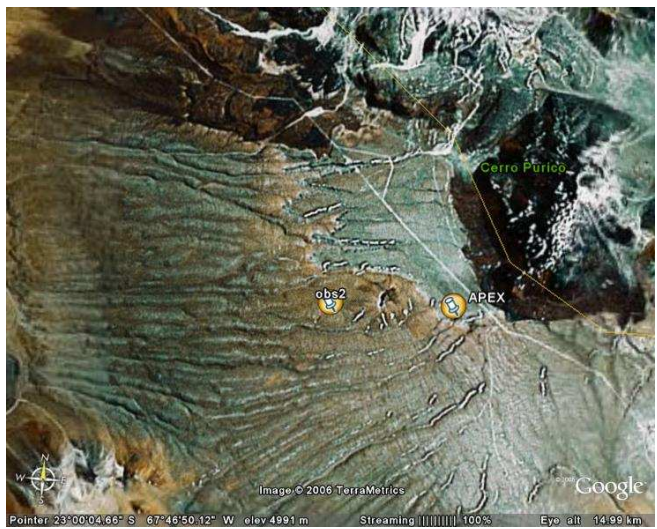

**Fig. 4.** Observing spot at $\sim$ 4950m. The approximate location of the APEX antenna at 5150m is also indicated

The first night, on 26 November, was exclusively used to position and align the telescope and the camera. The drift technique was used for alignment to the South Pole. With the current instrument and software used (Pegasus2000), the NS alignment is tricky and delicate. We completed the polar alignment within an error of $\leq$ 3 arcmin/30$^o$.

During the nights of 27 and 28 November, we monitored the luminosity of bright standard stars from zenith angle 0$^o$ to 60$^o$ (1 to 2 air masses), in U, B, & V (180 images collected) deducing this way the extinction coefficients. The following nights, we observed two SA95 fields one centered on HD 24622 which contains 5 Landolt standard stars (1992) and a second one centered on HD 24401 which contains 14 standard stars. These data were used to determine the color equations and the NSB in UBV. The response curves of the filters are shown in Fig. 2.

Figure 3 is a U band image of the field HD 24401 showing the Landolt stars (1992) used.

The six observing nights at La Silla were photometric. The observing logs for the fields HD 24622 and HD 24401 are given in Table 1.

At 4000-5000 m, the ALMA collaboration provided logistic and medical support, and a very friendly help. Thanks to this support we were able to continue our project. The observation camps were fixed at km 27.5 (4000m) and km 40.7 (4950m) respectively, along the ALMA road that leads from the Observing Support Facility (OSF) to the Chajnantor plateau. Our position at $\sim$ 4950m is shown in Fig. 4. The ambient temperature during the nights was around 0 C to -5 C. Since we were all coming from sea level, we had to face the question of night work combining low temperature, with low oxygen and high altitude. Our approach of working at 4000 m during 2-3 nights after 24 h at the OSF, and before going at 5000 m was appropriate. After one night at 5000 m we could all work well. The use of oxygen from time to time facilitated our tasks.

Each night started with a quick polar alignment sequence. Because of daily wind, the tube and instrument were dismounted by the end of each night, after acquiring 2-3 darks of 20 or 30 min, and of 3 min.

During our high altitude run we encountered variable conditions, with thin and thick cirrus, periods of high (frozen) humidity, and clouds. We have defined the following classifica-tion to compare our nights: (1): photometric all the night, no cirrus, no clouds on the horizon; (2): photometric with some haze compared with (1); (3): thin cirrus, photometric during an undefined period of the night; (4): thick cirrus; (5): high humidity, frost; (6): clouds. When we say that there are cirrus this means that there were cirrus in the late afternoon but we don't see them during the night.

**Table 1.** Observing log of the fields HD 24401 and HD 24622 (SA 95) at La Silla

| Night | Quality | Field | U | Air mass (U) | B | Air mass (B) | V | Air mass (V) |
|---|---|---|---|---|---|---|---|---|
| 28/11/2005 (-2.5*) | (1) | HD24622 | 1200s | 1.20 | 600s | 1.17 | 300s | 1.15 |
|  |  | HD24622 | 1200s | 1.51 | – | – | 300s | 1.28 |
| 29/11/2005 (-1.5*) | (1) | HD24622 | 1200s | 1.19 | 600s | 1.15 | – | – |
|  |  | HD24622 | 1200s | 1.17 | 600s | 1.44 | – | – |
|  |  | HD24622 | 1200s | 1.15 | 600s | 1.83 | – | – |
|  |  | HD24622 | 1200s | 1.18 | – | – | – | – |
|  |  | HD24622 | 1200s | 1.22 | – | – | – | – |
|  |  | HD24622 | 1200s | 1.29 | – | – | – | – |
|  |  | HD24401 | 1200s | 1.58 | 600s | 1.74 | 300s | 1.44 |
| 30/11/2005 (-0.5*) | (1) | HD24622 | – | – | 600s | 1.11 | 300s | 1.09 |
|  |  | HD24401 | 1200s | 1.26 | 600s | 1.14 | 300s | 1.18 |
|  |  | HD24401 | 1200s | 1.18 | 600s | 1.17 | 300s | 1.19 |
|  |  | HD24401 | 1200s | 1.15 | – | – | – | – |
|  |  | HD24401 | 1200s | 1.14 | – | – | – | – |
|  |  | HD24401 | 1200s | 1.20 | – | – | – | – |
| 01/12/2005 (0.5*) | (1) | HD24622 | 1200s | 1.20 | 600s | 1.17 | 300s | 1.16 |
|  |  | HD24401 | 1200s | 1.35 | 600s | 1.26 | 300s | 1.24 |

\* Number of days relative to new Moon

The night on the 4th of December had some cirrus. The beginning on the 5th was photometric but the moon became red when reaching the horizon level. The extinction clearly increased after 2 hours and became chaotic. None of nights of the end of March run were fully photometric all night. On the nights of March 24 & 25 there were some thin cirrus, but the reduced data indicate that the extinction curve was smooth ans the sky background in U was pretty stable. On March 26, the afternoon cirrus were thicker, and the data indicate more extinction. The night on March 27 was photometric. We obtained U band observations at 5000m during 2h 30. On the late afternoon of March 28 there were thin cirrus, but the extinction curve in U is the same as on March 27. Nevertheless there were frequent lightnings to the NE in the direction to Bolivia. We obtained U, B, V extinction curves and measurements of the NSB in U. The night on March 29 was quickly stopped due to frost. During the late afternoon on March 30 there were thick cirrus. The observations started after a period of high frozen humidity and haze. March 31 was cloudy.

To summarize we have acquired enough data at 4000 m to have an idea of the sky background and the extinction in U. At 5000m there were two partly photometric nights for which the extinction and sky background can be derived. There were some short periods of really dark sky background and high transparency.

The observing logs at 4000m and 5000m are given in Table 2 and Table 3 respectively.

## 3. Data and Analysis

The data were recorded in FITS format. Header information includes 39 parameters coming from the CCD camera, the filter wheel and the telescope. The CCD temperature was kept at -24.0 C with a precision of 0.1 C. If the ambient temperature was high, the CCD temperature was kept at -23.55 C so that the Peltier cooling was used only at 80%. The gain of the CCD was 2.65. Saturation level occurs at ~ 100000 $e^-$. Dark current is estimated at 1 $e^-$/pixel/sec at 0 C. With this type of detector we are obliged to acquire several dark current runs at the same temperature. The noise measured on a dark exposure contains two contributions, one is the variation of pedestals which is reproducible and dominates, and a second which is random is true not-reproducible noise. The readout noise is 15 $e^-$ rms. The variation of pedestals is considerably increased if a 2x2 binning is used. To give an example, the noise measured for a dark of 1x1 binning and integration time of 5 min is 37 ADU rms, while for a dark of 20 min with 2x2 binning we measured 253 ADU rms. After subtractions of 2 darks of 20 min binned 2x2, we achieved a noise of 12 ADU rms. For each image recorded we subtracted an average dark of the same exposure time. After dark subtraction, the sky luminosity measured on a 10 min exposure in B was typically 75 ADU with a noise of 12 ADU rms, while on 20 min exposures with 2x2 binning in U it was 50 ADU with noise level of 14 ADU rms. The shutter is electromagnetic with a resolution of 10 ms on the range between 0.12 s and 3600 s. Master-flats were built by median filtering of the NSB exposures acquired during the runs.

### 3.1. La Silla Observations

The extinction was measured during the 27 and 28 November 2005 nights by observing the same bright stars at air masses $1 \leq Z \leq 2$ in U, B, & V. We used exposure sequences of U (60s) - Dark (60s) - B (20s) - Dark (20s) - V (20s) - Dark (20s). The dark subtraction was done immediately after exposure and the difference only was recorded. The Bouguer curves in U, B,

**Table 2.** Observing log at 4000 m on ALMA site (fields SA 95, SA 101 and SA 104)

| Night | Quality | Field | U (in s) | Air mass (U) | B | Air mass (B) | V | Airmass(V) |
|---|---|---|---|---|---|---|---|---|
| 04/12/2005 (3.5*) | 3 | HD24622 | 24 × 180 | 1.213 - 1.877 | | | | |
| 05/12/2005 (4.5*) | 3 | HD24622 | 1200 | 1.13 | | | | |
| | | HD24622 | 1200 | 1.09 | | | | |
| | | HD24622 | 15 × 180 | 1.273 - 1.453 | | | | |
| 24/03/2006 (-4.3*) | 3 | HD86135 | 35 × 180 | 1.084 - 1.89 | | | | |
| 25/03/2006 (-3.3*) | 3 | SA101421 | 1800 | 1.10 | | | | |
| | | SA101421 | 1800 | 1.09 | | | | |
| | | SA101421 | 1800 | 1.09 | 600 | 1.13 | 300 | 1.12 |
| | | SA104456 | 1800 | 1.09 | | | | |
| | | SA104456 | 1800 | 1.13 | | | | |
| 26/03/2006 (-2.3*) | 4 | HD86135 | 38 × 180 | 1.084 - 2.32 | | | | |
| | | SA101421 | 1800 | 1.14 | | | | |

**Table 3.** Observing log at 5000 m on ALMA site (fields SA 101 and SA 104)

| Night | Quality | Field | U (in s) | Air mass (U) | B | Air mass (B) | V | Airmass(V) |
|---|---|---|---|---|---|---|---|---|
| 27/03/2006 (-2.3*) | 1 | HD86135 | 24 × 180 | 1.085 - 1.47 | | | | |
| 28/03/2006 (-1.3*) | 3 | HD86135 | 25 × 180 | 1.09 - 2.11 | 25 × 60 | 1.10 - 2.13 | 25 × 10 | 1.09 - 2.04 |
| | | SA101421 | 1800 | 1.13 | | | | |
| | | SA101421 | 1800 | 1.09 | | | | |
| | | SA101421 | 1800 | 1.30 | | | | |
| 30/03/2006 (0.7*) | 4 | SA101421 | 1800 | 1.34 | | | | |
| | | SA104456 | 1800 | 1.09 | | | | |

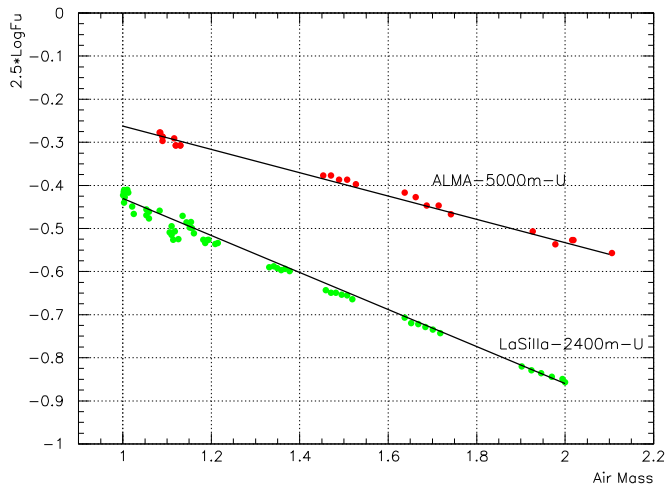

**Fig. 5.** Bouguer curve in $U_{obs}$ for the nights on 27 and 28 November 2005 on La Silla and March 28 on Chajnantor. The graphic shows the luminosity variation $2.5 \times [\text{Log}(F_U)_{Z=0} - \text{Log}(F_U)(Z)]$ as a function of the airmass for the U filter, for each site

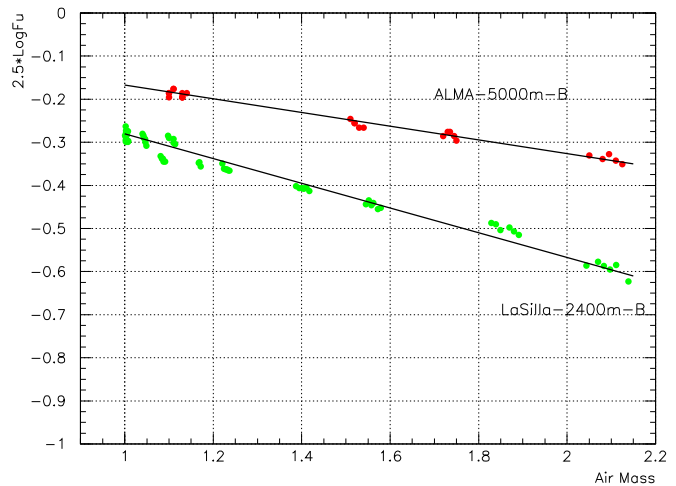

**Fig. 6.** Bouguer curves in $B_{obs}$ for the same nights as in Fig. 5

V (observed) are shown in Fig. 5, 6, 7. The flux measurements and the slopes of the extinction curves are the same for the two nights. The slopes are:

Extinction ($U_{obs}$) = 0.424 ± 0.01 mag/airmass
Extinction ($B_{obs}$) = 0.271 ± 0.01 mag/airmass
Extinction ($V_{obs}$) = 0.164 ± 0.01 mag/airmass

The color equations and the NSB were derived from Landolt standard stars measured in the HD 24401 and HD 24622 fields during the nights on 28-30 November and December 1rst.

The exposure times were 300 s in V and 600 s in B with a pixel of 1.7 arcsec, and 1200 s in U with a pixel of 3.4 arcsec.

The errors were estimated from large scale variations over the CCD (< 3 %), from local noise, and from differences of measurements of a same object. For 1200 s exposures and a 2x2 binning, average errors in $U_{obs}$ are 0.02 mag for U ≤ 14, 0.05 mag for U ≈ 15 et 0.1 mag for U ≈ 16.5.

We obtained the following equations (Figs. 8, 9, 10) between observed and standard magnitudes, and colors:

$U = A_U(t) + (0.27 \pm 0.07) \times (U - B) - 2.5 \times \text{Log}F(U_{obs})$
$B = A_B(t) + (0.14 \pm 0.03) \times (B - V) - 2.5 \times \text{Log}F(B_{obs})$
$V = A_V(t) + (-0.12 \pm 0.03) \times (B - V) - 2.5 \times \text{Log}F(V_{obs})$

### 3.1.1. Night sky background in U

The values of the NSB in U are given in Table 4. We have indicated the instrumental surface brightness $2.5 \times \text{Log}F(U_{obs})$ per arcsec measured on CCD surfaces including more than 1000 pixels. The average is of the order of 50 ADU per pixel with a noise of 12 ADU rms. The magnitude background has been derived from the observed flux, corrected for extinction and $U - B$ color for the standard stars. The error in sky background luminosity is dominated by differences between standard stars. The sky background in the HD 24401 field is derived from 9 stars. For the HD 24622 field (4 stars in U) there is a systematic difference of 0.25 mag in $A_U$ between the stars SA 95132 et SA 95139. We have assumed that the true color of the sky is $U - B = -0.70$.

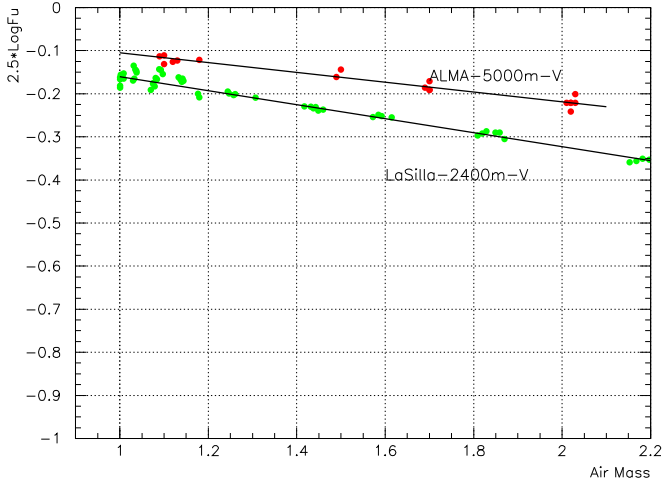

**Fig. 7.** Bouguer curves in $V_{obs}$ for the same nights as in Fig. 5

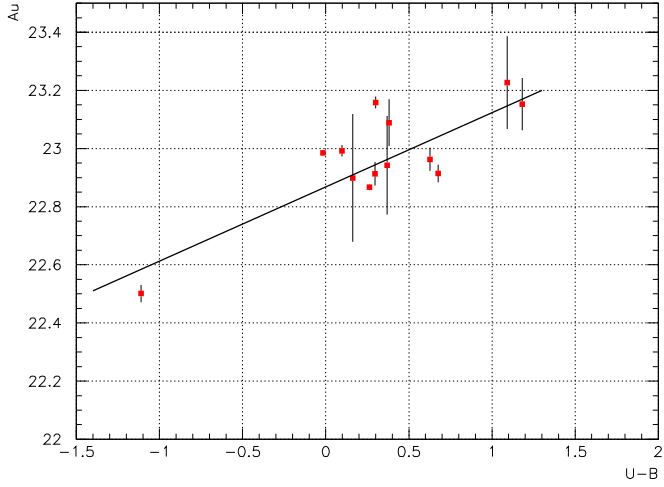

**Fig. 8.** Color equation between calibrated colors $U - B$ from Landolt and $A_U = 2.5 \times \mathrm{LogF}(U_{obs}) + U$. $A_U$ depends on the exposure time. The equations were obtained after correction for differences in air masses. Each point is the average of 5 or more measurements.

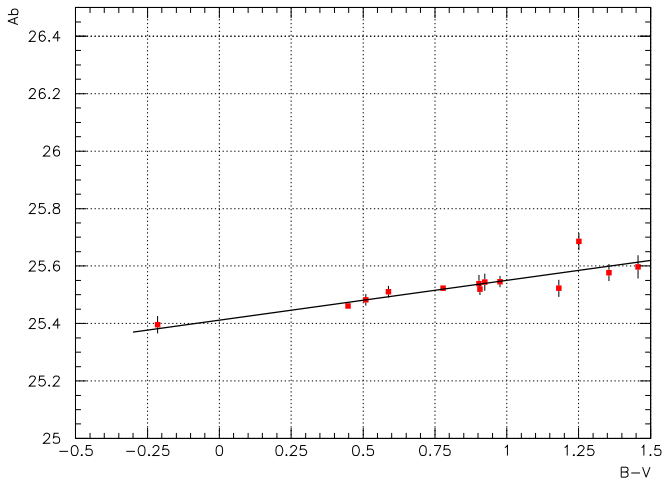

**Fig. 9.** Color equation between calibrated colors $B - V$ from Landolt and $A_B = 2.5 \times \mathrm{LogF}(B_{obs}) + B$.

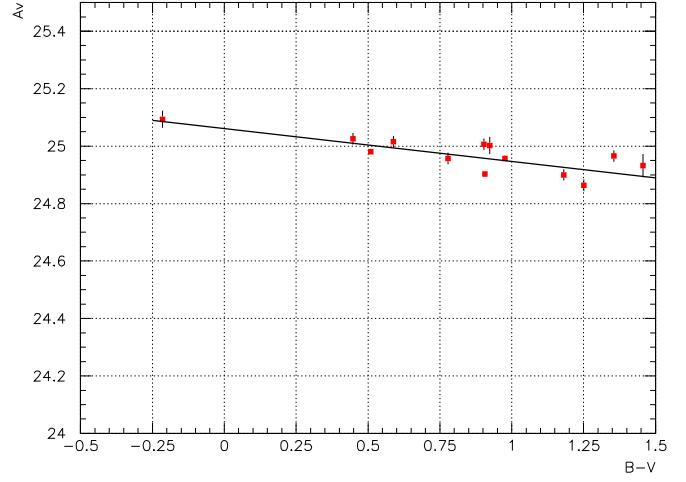

**Fig. 10.** Color equation between calibrated colors $B - V$ from Landolt and $A_V = 2.5 \times \mathrm{LogF}(V_{obs}) + V$.

**Table 4.** Sky background in U at La Silla

| Night | Field | $2.5 \times \mathrm{LogF}(U_{obs})$ | NSB in U |
|---|---|---|---|
| 28 Nov (-2.5*) | HD24622 | 1.71 ± 0.02 | 22.01 ± 0.09 |
|  | HD24622** | 1.70 ± 0.02 | 22.05 ± 0.09 |
| 29 Nov (-1.5*) | HD24622 | 1.67 ± 0.02 | 22.09 ± 0.12 |
|  | HD24622 | 1.72 ± 0.02 | 22.04 ± 0.12 |
|  | HD24622** | 1.59 ± 0.02 | 22.07 ± 0.09 |
|  | HD24622 | 1.58 ± 0.02 | 22.08 ± 0.09 |
|  | HD24622** | 1.57 ± 0.02 | 22.09 ± 0.09 |
|  | HD24622** | 1.57 ± 0.02 | 22.09 ± 0.09 |
|  | HD24401 | 1.55 ± 0.02 | 22.11 ± 0.04 |
| 30 Nov (-0.5*) | HD24401 | 1.62 ± 0.02 | 21.96 ± 0.05 |
|  | HD24401 | 1.62 ± 0.02 | 21.97 ± 0.03 |
|  | HD24401 | 1.61 ± 0.02 | 21.95 ± 0.03 |
|  | HD24401 | 1.45 ± 0.02 | 22.20 ± 0.03 |
|  | HD24401 | 1.59 ± 0.02 | 21.99 ± 0.03 |
| 1 Dec (0.5*) | HD24622 | 1.67 ± 0.02 | 22.01 ± 0.09 |
|  | HD24401 | 1.61 ± 0.02 | 22.00 ± 0.03 |

### 3.1.2. Night sky background in B

The values of the NSB in B are given in Table 5. We have indicated the instrumental surface brightness $2.5 \times \mathrm{LogF}(B_{obs})$ per arcsec measured on CCD surfaces including more than 1000 pixels. The average is of the order of 80 ADU per pixel with a noise of 12 ADU rms. The NSB luminosity has been derived from the observed flux, corrected for extinction and $B - V$ color for the standard stars.

**Table 5.** Sky background in B at La Silla

| Night | Field | $2.5 \times \text{Log} F(B_{obs})$ | NSB in B |
|---|---|---|---|
| 28 Nov (-2.5*) | HD24622 | 3.53 ± 0.02 | 22.73 ± 0.05 |
| 29 Nov (-1.5*) | HD24622 | 3.51 ± 0.02 | 22.74 ± 0.05 |
|  | HD24622 | 3.48 ± 0.02 | 22.79 ± 0.05 |
|  | HD24622 | 3.49 ± 0.02 | 22.81 ± 0.05 |
|  | HD24401 | 3.44 ± 0.02 | 22.84 ± 0.03 |
| 30 Nov (-0.5*) | HD24401 | 3.57 ± 0.02 | 22.65 ± 0.03 |
|  | HD24401 | 3.61 ± 0.02 | 22.60 ± 0.03 |
|  | HD24401 | 3.61 ± 0.02 | 22.66 ± 0.03 |
| 1 Dec (0.5*) | HD24622 | 3.53 ± 0.02 | 22.71 ± 0.05 |
|  | HD24401 | 3.50 ± 0.02 | 22.70 ± 0.03 |

**Table 6.** Sky background in V at La Silla

| Night | Field | $2.5 \times \text{Log} F(V_{obs})$ | NSB in V |
|---|---|---|---|
| 28 Nov (-2.5*) | HD24622 | 3.73 ± 0.02 | 21.68 ± 0.05 |
|  | HD24622 | 3.62 ± 0.02 | 21.79 ± 0.05 |
| 29 Nov (-1.5*) | HD24401 | 3.48 ± 0.02 | 21.91 ± 0.03 |
| 30 Nov (-0.5*) | HD24401 | 3.68 ± 0.02 | 21.71 ± 0.03 |
|  | HD24401 | 3.66 ± 0.02 | 21.73 ± 0.03 |
|  | HD24622 | 3.58 ± 0.02 | 21.81 ± 0.05 |
| 1 Dec (0.5*) | HD24622 | 3.60 ± 0.02 | 21.78 ± 0.05 |
|  | HD24401 | 3.62 ± 0.02 | 21.76 ± 0.03 |

### 3.1.3. Night sky background in V

The values of the NSB in V are given in Table 6. We have indicated the instrumental surface brightness $2.5 \times \text{Log} F(V_{obs})$ per arcsec measured on CCD surfaces including more than 1000 pixels. The NSB luminosity has been derived from the observed flux, corrected for extinction and $B - V$ color for the standard stars.

### 3.2. Observations at 4000m asl

#### 3.2.1. December 2005 run

Series of 180 s exposures of the field HD 24622 were obtained in automatic mode. This mode allowed us to measure the extinction as a function of air mass from the stars HD 24622 and SA 95132, and to obtain images with integrated exposure time similar to La Silla, by co-adding groups of 7 images. The sky in the late afternoon of December 4 showed some cirrus (Fig. 12). The Bouguer curve in U for the night on December 4 is shown in Fig. 13. The slope is:

Extinction ($U_{obs}$) = 0.33 ± 0.02 mag/airmass

which is significantly more transparent than on La Silla. Another way to measure the difference in extinction 4000 m - La Silla is to directly compare the instrumental fluxes of the same stars at the same air masses. We have built 3 images of 20 min with average air masses 1.20, 1.32 and 1.53 from the data at 4000 m which we have compared to images of the same air masses obtained on La Silla (marked by ** in the Table 4. The extinction difference from these images is:

Z = 1.20: $\triangle U_{obs}$ = 0.09 mag,
Z = 1.32: $\triangle U_{obs}$ = 0.12 mag,
Z = 1.53: $\triangle U_{obs}$ = 0.14 mag,

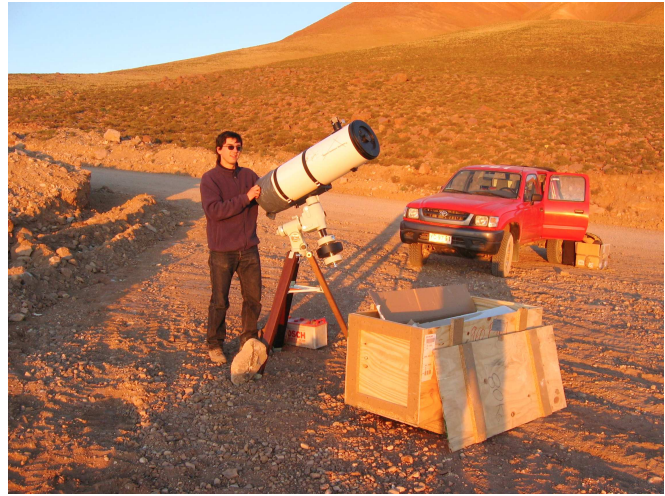

**Fig. 11.** Mounting the telescope at 4000m near the ALMA road, in the late afternoon on March 24

The beginning of the night on December 5 is similar to the 4th (a few cirrus) but the sky changes after 2 hours of observations. The extinction increases and fluctuates, and the sky becomes brighter. At the beginning of the night (first exposure of 1200s in Table 2.):

Z = 1.13: $\triangle U_{obs}$ = 0.08 mag

#### 3.2.2. March 2006 run

We used the same spot as on December (Fig.11). Some thin cirrus, which were visible during the day, may have been present during parts of the nights on March 24 and 25. Nevertheless there is no significant fluctuation in the Bouguer curve on March 24. We classified the sky quality of these nights as "photometric during an undefined period of the night". We observed 3 fields. They were centered on HD 86135 in SA 101, on SA 101421, and on SA 104456 respectively.

On March 24, we obtained series of 180 s exposures in U on HD 86135, to measure the extinction, and to derive the sky background by co-adding images. In order to measure the sky background with lower noise, we acquired 1800s U-band exposures of the fields SA 101 and SA 104 on March 25.

The night on March 26 was not photometric. We did the same monitoring as on March 24. The Bouguer curve is found to be convex (more extinction before 1:00 local time than after). This night is rejected.

The slope of the extinction curve in U on March 24 was

Extinction ($U_{obs}$) = 0.37±0.02 mag/airmass measured over

the range 1.085 ≤ Z ≤ 1.82.

We have indicated in Table 7 the values of the sky background in U. The instrumental magnitudes are given for an exposure time of 20 min for comparison, whenever they were derived from 30 min exposures or from co-added images of 180s. The sky background is derived for an extinction of $U_{obs}$ = 0.33 mag/airmass on December 4-5 and $U_{obs}$ = 0.37 mag/airmass on March 24-25. For the instrumental color correction, we have assumed a true sky background color of $U - B = -0.7$, like on La Silla.

**Table 7.** Sky background in U at 4000 m

| Night | Field | $2.5 \times LogF(U_{obs})$ | NSB in U |
|---|---|---|---|
| 04/12/2005 (+3.5*) | HD24622 | 1.52 ± 0.03 | 22.16 ± 0.09 |
| | HD24622 | 1.41 ± 0.03 | 22.27 ± 0.09 |
| | HD24622 | 1.48 ± 0.03 | 22.17 ± 0.09 |
| 05/12/2005 (+4.5*) | HD24622 | 1.53 ± 0.02 | 22.09 ± 0.09 |
| | HD24622 | 1.64 ± 0.02 | 21.94 ± 0.09 |
| | HD24622 | 1.99 ± 0.02 | – |
| | HD24622 | 1.95 ± 0.02 | – |
| 24/03/2006 (-4.3*) | HD86135 | 1.54 ± 0.03 | 22.14 ± 0.09 |
| | HD86135 | 1.51 ± 0.03 | 22.10 ± 0.09 |
| | HD86135 | 1.36 ± 0.03 | 22.27 ± 0.09 |
| 25/03/2006 (-3.3*) | SA101421 | 1.49 ± 0.02 | 22.12 ± 0.05 |
| | SA101421 | 1.52 ± 0.02 | 22.10 ± 0.05 |
| | SA101421 | 1.48 ± 0.02 | – |
| | SA104456 | 1.31 ± 0.02 | 22.19 ± 0.07 |
| | SA104456 | 1.42 ± 0.02 | 22.06 ± 0.07 |

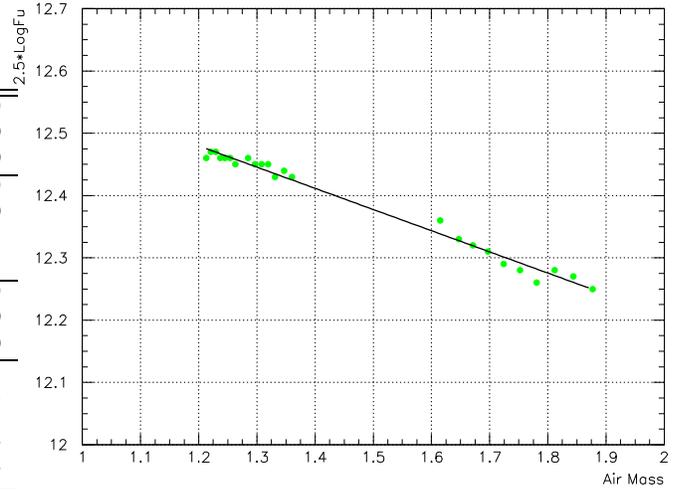

**Fig. 13.** Bouguer curve in $U_{obs}$ for the night on 4 December 2005 at 4000m asl. The graphic shows the observed luminosity variation 2.5 Log($F_U$), in arbitrary units, as a function of air mass

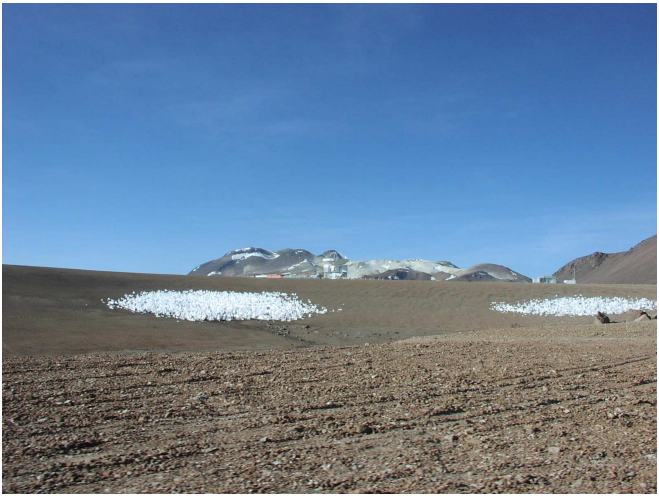

**Fig. 12.** The sky in the late afternoon of December 4, on the Chajnantor plateau showing some cirrus.

We find that in spite of a few possible cirrus the nights on December 4 and March 24 were more transparent and darker in U than our photometric nights on La Silla. The beginning of the night on December 5 is of the same quality. The night on March 25 was less homogeneous. In particular there was an additional extinction of 0.15 mag in the third U-band 30 min exposure on SA 101421, compared with the same field exposure 1h before, but the instrumental sky background had not changed. With a correction of 0.15 mag the NSB measured on this frame would be U = 22.13. Two hours later the sky was again transparent and very dark (next exposure on SA 104456 in Table 7). There is nothing peculiar on the NSB of our single B and V images: the values of the surface brightnesses are B = 22.70 ± 0.03 mag arcsec$^{-2}$ V = 21.64 ± 0.03 mag arcsec$^{-2}$ respectively.

Since we were never convinced that we were observing during nights of excellent photometric quality, it seems very reasonable to expect a U-band sky background of at least 22.10 – 22.15 mag arcsec$^{-2}$ in a photometric dark night.

### 3.3. Observations at 5000m asl

The night on March 27 was a test night. It was photometric. After doing a polar alignment of the telescope, we obtained a series of 180s exposures in automatic mode of the field HD 86135 to measure the extinction in U. We stopped observing at 1:am. The sky background in U was derived from two images, 21 min each, obtained by coadding two sets of 7 images, 180s each, at average air-masses Z = 1.10 and Z = 1.22 respectively. The measured NSB are U = 22.24 ± 0.09 mag arcsec$^{-2}$ and U = 22.32 ± 0.09 mag arcsec$^{-2}$ (Table 8).

Some thin cirrus, which were visible during the day on March 28, may have been present during part of the night. Frequent lightnings were visible to the NE on Bolivia. We started that night with a 30 min U-band exposure at Z = 1.14 before meridian, on which we measure a NSB of U = 22.37 ± 0.05 mag arcsec$^{-2}$ (Fig. 14), then we made a series of 180s exposures for measuring the extinction at low airmass. Coadding 10 images at Z = 1.09-1.13 yields U = 22.39 ± 0.09 mag arcsec$^{-2}$. The next 30 min exposure in U was obtained at Z = 1.15 after meridian. Both instrumental and calibrated measurements indicate that the sky had become brighter: U = 22.10 ± 0.05 mag arcsec$^{-2}$. Later on, a 30 min exposure at Z = 1.30 gave U = 21.93 ± 0.05 mag arcsec$^{-2}$. We have coadded 10 images of 60s in B with low airmass: they give a surface brightness of B = 22.87 ± 0.05 mag arcsec$^{-2}$.

The Bouguer curves in U, B, V for that night are shown in Figs. 5, 6, 7 together with those obtained on La Silla.

The slopes are: .

Extinction ($U_{obs}$) = 0.26 ± 0.01 mag/airmass
Extinction ($B_{obs}$) = 0.16 ± 0.01 mag/airmass
Extinction ($V_{obs}$) = 0.11 ± 0.01 mag/airmass

The same U-band extinction curve was found on March 27, in the shorter range Z = 1.09 to 1.47.

On the night of March 29, frost grew on the tube soon after mounting it. On March 30, there were thick cirrus in the afternoon (see Fig. 15). We mounted the tube after the frozen humidity decreased. On that night the measured background was higher, U = 21.97 − 21.91 mag arcsec$^{-2}$, but very similar to La Silla under photometric conditions.

**Table 8.** Sky background in U at 5000m

| Night | Field | $2.5 \times \mathrm{LogF(U_{obs})}$ | NSB in U |
|---|---|---|---|
| 27/03/2006 (-2.3*) | HD86135 | 1.30 ± 0.04 | 22.24 ± 0.09 |
| | HD86135 | 1.23 ± 0.04 | 22.32 ± 0.09 |
| 28/03/2006 (-1.3*) | HD96135 | 1.22 ± 0.04 | 22.39 ± 0.09 |
| | SA101421 | 1.33 ± 0.02 | 22.37 ± 0.05 |
| | SA101421 | 1.53 ± 0.02 | 22.10 ± 0.05 |
| | SA101421 | 1.62 ± 0.02 | 21.93 ± 0.05 |
| 30/03/2006 (0.7*) | SA101421 | 1.62 ± 0.02 | – |
| | SA101421 | 1.66 ± 0.02 | – |

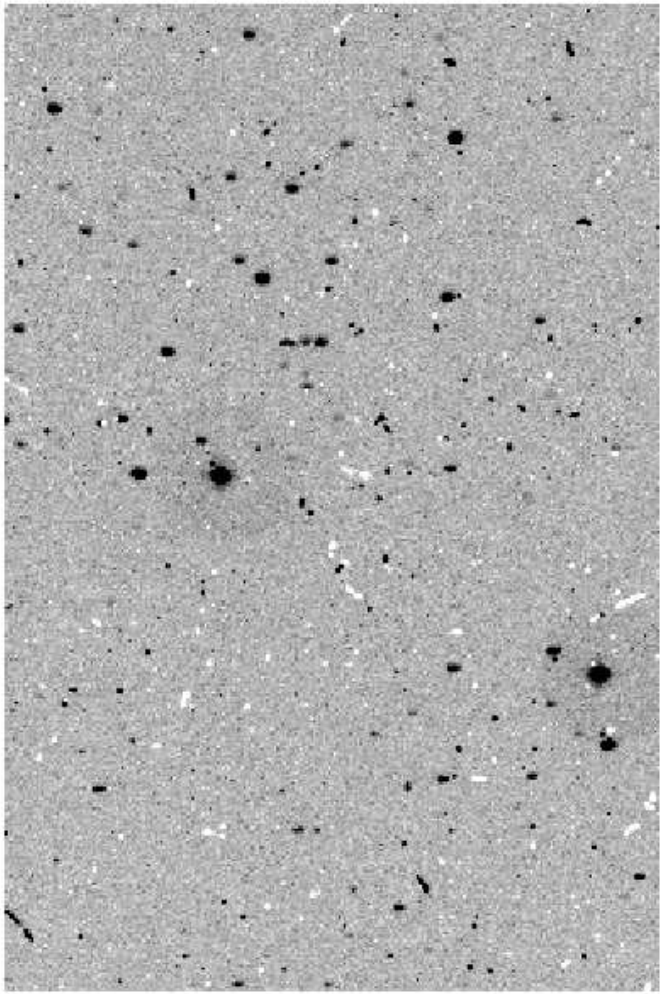

**Fig. 14.** U-band 1800s exposure at 5000m on which we measured a sky background of 22.37 ± 0.05 mag arcsec$^{-2}$. A large number of cosmic rays are visible on the frame and on the negative dark. North is up.

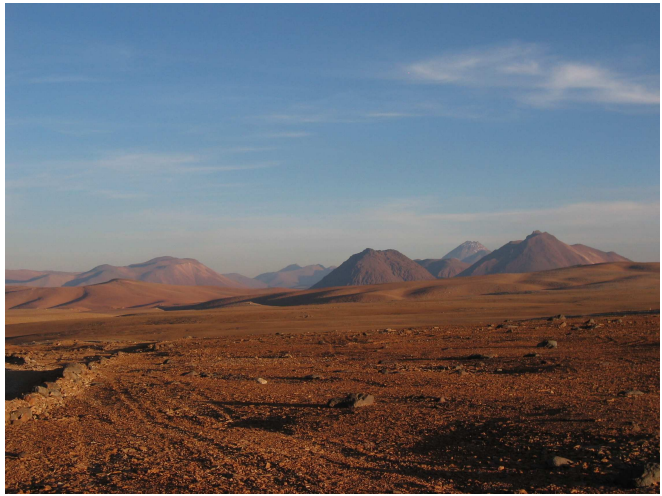

**Fig. 15.** The sky in the late afternoon of March 30 showing thick cirrus.

Our run is not conclusive concerning the background because we did not collect enough data in photometric conditions. Nevertheless we measured a NSB of U = 22.28 mag arcsec$^{-2}$ during 42 min on March 27 and U = 22.38 mag arcsec$^{-2}$ during 1 h on March 28.

Considering the sky conditions we had, one may suspect a significant variation of sky background with altitude rather than a fluctuation, but this still needs to be confirmed by a week of photometric measurements in a stable period.

## 4. Conclusions

We have observed the sky in U, B, & V during dark photometric nights on La Silla. Our extinction values, obtained by observing bright stars from Z = 1 to 2, during the nights on 27 and 28 November 2005 are similar to the average values recorded at La Silla (1993), but we did not compare with other telescopes during the same nights.

Our La Silla measurements suggest that the NSB in U was slightly darker, by 0.14 mag during the dark nights from 28 November to 1 December, than the average values obtained by monitoring over long periods (La Silla 2002).

The sky in U at 4000m, near Chajnantor, was more transparent by 0.08 mag/airmass than our photometric measurements on La Silla obtained with the same instrument. The extinction coefficient in U is close to the value given in the literature for Mauna Kea (2001, 2005). The sky background was also found somewhat darker in U, by 0.11 mag, than the values which we obtained on La Silla, and 0.25 mag fainter than the average value obtained over long periods of monitoring on La Silla.

The extinction in U, B, V at 5000m, on Chajnantor, was found ∼ 2/3 that on La Silla. Our 5000m run is not conclusive concerning the NSB but we could measure a clean dark sky background during some brief periods: on March 28, during 1h, we measured a ratio (star flux /background flux) that was 60 % higher than on La Silla.

Combining the data at 4000m and 5000m, the NSB in U seems to have a significant variation with altitude, compared with 2500m. It needs to be confirmed by a week of photometric measurements in a stable period at 5000m, and further monitored during long periods with an automatic system. Measurements of the NSB in B and V need to be done.

*Acknowledgements.* The telescope and camera were paid by the University of Montpellier through a grant from the IPM, and the cam-paign was supported by LPTA, with local support from ESO and ALMA. I. Toledo was partially supported by the FONDAP Centro de Astrofsica. Our thanks go in particular to M. Tarenghi, J. Lassalle, and E. Donoso (ALMA), M. Sarazin, and J. Spyromilio (ESO), E. Hardy (NRAO). We thank A. Falvard for is constant support, J. Lascaray (IPM), B. Gil (GES) and H. Quintana (PUC).